\documentclass[twoside]{ilcws08}
\usepackage[latin1]{inputenc}
\usepackage[dvips]{graphicx,epsfig,color}
\usepackage{wrapfig,rotating}
\usepackage{amssymb,amsmath,array}

\newcommand{\lhhh}{$\lambda_{HHH}$}
\newcommand{\eezhh}{$e^{+} e^{-} \to ZHH$}
\newcommand{\zhhnnhh}{$ZHH \to \nu \bar{\nu} HH$}
\newcommand{\nnhh}{$\nu \nu HH$}
\begin{document}
\title{Analysis of $ZHH$ in the 4-jet mode} %% 
%***********************************************************************
% AUTHORS INFORMATION AREA
%***********************************************************************
\author{Yosuke Takubo
%$^1$ and Second Author$^2$
% Optional short acknowledgment: remove next line if non-needed
%\thanks{This is an optional funding source acknowledgment.}
% DO NOT MODIFY THE FOLLOWING '\vspace' ARGUMENT
\vspace{.3cm}\\
% Addresses and institutions (remove "1- " in case of a single institution)
Department of Physics, Tohoku University, Sendai, Japan
%% Remove the next three lines in case of a single institution
%\vspace{.1cm}\\
%2- Second Author's Institution - Department \\
%Address of Second Author's Institution - Country\\
}
%%***********************************************************************
% END OF AUTHORS INFORMATION AREA
%***********************************************************************

\maketitle

\begin{abstract}
Measurement of the cross-section of \eezhh~offers the information of
the trilinear Higgs self-coupling,
which is important to confirm the mechanism of the electro-weak symmetry breaking.
Since there is huge background in the signal region,
background rejection is key point to identify $ZHH$ events.
In this paper, we study the possibility to observe the $ZHH$ events at ILC 
by using \zhhnnhh, resulting 4 jets.
\end{abstract}

%%%%%%%%%%%%%%%%%%%%%%%%%%%%%%%%%%
\section{Introduction}
%%%%%%%%%%%%%%%%%%%%%%%%%%%%%%%%%%
In the standard model,
particle masses are generated through the Higgs mechanism.
This mechanism relies on a Higgs potential, 
$V(\Phi) = \lambda (\Phi^{2} - \frac{1}{2} v^{2})^{2}$,
where $\phi$ is an iso-doublet scalar field, and $v$  
is the vacuum expectation value of its neutral component ($v \sim 246$ GeV).
Determination of the Higgs boson mass,
which satisfies $m_{H}^{2} = 2 \lambda v^{2}$ at tree level in the standard model,
will provide an indirect information about the Higgs potential and its self-coupling, \lhhh.
The measurement of the trilinear self-coupling, 
$\lambda_{HHH} = 6 \lambda v$, offers 
an independent determination of the Higgs potential shape
and the most decisive test of the mechanism of the electro-weak symmetry breaking.

\begin{wrapfigure}{r}{0.5\columnwidth}
\centerline{\includegraphics[width=0.45\columnwidth]{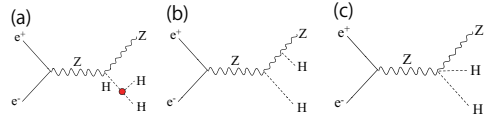}}
\caption{The relevant Feynman diagrams for the $ZHH$ production. 
The trilinear self-coupling is included in (a).}
\label{fig:diagram}
\end{wrapfigure}

\lhhh~can be extracted from the measurement of the cross-section 
for the Higgs-strahlung process ($\sigma_{ZHH}$), \eezhh.
For a Higgs mass of 120 GeV, the $W$ fusion process is negligible
at $\sqrt{s} = 500$ GeV.
Figure \ref{fig:diagram} shows the relevant Feynman diagrams for this process.
The information of \lhhh~is included in the diagram of Fig. \ref{fig:diagram}(a),
and the relation between the cross-section of 
$ZHH$ and \lhhh~is characterized by
$\frac{\Delta \lambda_{HHH}}{\lambda_{HHH}} 
\sim 1.75 \frac{\Delta \sigma_{ZHH}}{\sigma_{ZHH}}$,
where $\Delta \lambda_{HHH}$ and $\Delta \sigma_{ZHH}$
are measurement accuracy of $\lambda_{HHH}$ and $\sigma_{ZHH}$, respectively
\cite{castanier}.
For that reason, precise measurement of the cross-section for the $ZHH$ production
is essential to determination of the strength of the trilinear Higgs self-coupling.
In this paper, we report status of our feasibility study for
the observation of $ZHH$ events at the ILC.

%%%%%%%%%%%%%%%%%%%%%%%%%%%%%%%%%%
\section{Target mode}
%%%%%%%%%%%%%%%%%%%%%%%%%%%%%%%%%%
In this study, we assumed a Higgs mass of 120 GeV, 
$\sqrt{s} = 500$ GeV, and an integrated luminosity of 2 $ab^{-1}$.
The final states of the $ZHH$ production can be categorized into 3 types, 
depending on the decay modes of $Z$:
$ZHH \to q\bar{q} HH$ (135.2 ab$^{-1}$), 
$ZHH \to \nu\bar{\nu} HH$ (38.8 ab$^{-1}$), and
$ZHH \to \ell\bar{\ell} HH$ (19.8 ab$^{-1}$),
where the cross-sections were calculated 
without the beam polarization, initial-state radiation, and beamstrahlung.
Since $ZHH \to q\bar{q} HH$ has the largest cross-section,
this mode is the most attractive to observe $ZHH$ events.
However, we study \zhhnnhh~in this paper, 
because the 4-jet final state is easy to analyze.
For the Higgs mass of 120 GeV, the Higgs boson mainly decays into $b\bar{b}$ (76\% branching ratio).
Therefore, we concentrated on $ZHH \to \nu\bar{\nu} b\bar{b}b\bar{b}$
from \nnhh~events.
As background events, we considered $ZZ \to bbbb$ (9.05 fb), 
$tt$ (583.6 fb), $ZH$ (62.1 fb), and $tbtb$ (1.2 fb).
They have much larger cross-sections than $ZHH$,
necessitating powerful background rejection.

%%%%%%%%%%%%%%%%%%%%%%%%%%%%%%%%%%
\section{Simulation tools}
%%%%%%%%%%%%%%%%%%%%%%%%%%%%%%%%%%
\begin{wrapfigure}{r}{0.35\columnwidth}
\centerline{\includegraphics[width=0.3\columnwidth]{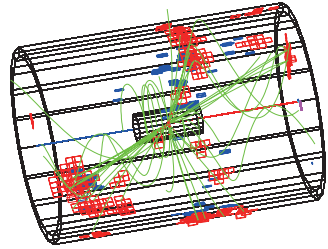}}
\caption{A typical event display of $ZHH \to \nu_{\mu} \bar{\nu}_{\mu} HH$.}
\label{fig:evtdsp}
\end{wrapfigure}

We have used MadGraph \cite{madgraph} 
to generate \nnhh~and $tbtb$ events, 
while $ZZ \to bbbb$, $tt$, and $ZH$ events have been generated 
by Physsim \cite{physsim}.
In this study, the beam polarization, initial-state radiation, and beamstrahlung 
have not been included in the event generations. 
We have ignored the finite crossing angle between the electron and positron beams. 
In both event generations, 
helicity amplitudes were calculated using the HELAS library \cite{helas}, 
which allows us to deal with the effect of gauge boson polarizations properly. 
Phase space integration and the generation of parton 4-momenta have been performed 
by BASES/SPRING \cite{bases}. 
Parton showering and hadronization have been carried out by using PYTHIA6.4 \cite{pythia}, 
where final-state tau leptons are decayed by TAUOLA \cite{tauola} 
in order to handle their polarizations correctly.

The generated Monte Carlo events have been passed to a detector simulator 
called JSFQuickSimulator, 
which implements the GLD geometry and other detector-performance 
related parameters \cite{glddod}. 
Figure \ref{fig:evtdsp} shows a typical event display of 
$ZHH \to \nu_{\mu} \bar{\nu}_{\mu} HH$.
%In the detector simulator, hits by charged particles at the vertex detector 
%and track parameters at the central tracker are smeared according 
%to their position resolutions. 
%Since calorimeter signals are simulated in individual segments, 
%a realistic simulation of cluster overlapping is possible. 
%Track-cluster matching is performed for the hit clusters in the calorimeter 
%in order to achieve the best energy flow measurements. 

%%%%%%%%%%%%%%%%%%%%%%%%%%%%%%%%%%
\section{Analysis}
%%%%%%%%%%%%%%%%%%%%%%%%%%%%%%%%%%
\begin{wrapfigure}{r}{0.38\columnwidth}
\centerline{\includegraphics[width=0.34\columnwidth]{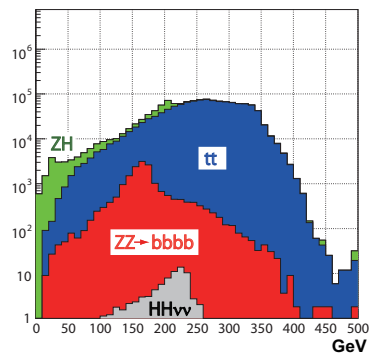}}
\caption{Distribution of the sum of the two reconstructed Higgs masses 
for \nnhh~and background events.}
\label{fig:nocut}
\end{wrapfigure}

The clusters in the calorimeters are combined to form a jet 
if the two clusters satisfy $y_{ij} < y_{\mathrm{cut}}$,
where $y_{ij}$ is $y$-value of the two clusters. 
All events are forced to have four jets by adjusting $y_{\mathrm{cut}}$.
Then, mass of the Higgs boson was reconstructed to identify \nnhh~events
by minimizing $\chi^{2}$ value defined as
\begin{equation}
\chi^{2} = 
\frac{(^{\mathrm{rec}}M_{H1} - ^{\mathrm{true}}M_{H})^{2}}{\sigma_{H1}^{2}} +
\frac{(^{\mathrm{rec}}M_{H2} - ^{\mathrm{true}}M_{H})^{2}}{\sigma_{H2}^{2}},
\end{equation}
where $^{\mathrm{rec}}M_{H1,2}$, $^{\mathrm{true}}M_{H1,2}$, and
$\sigma_{H1,2}$ are the reconstructed Higgs mass, the true Higgs mass (120 GeV),
and the Higgs mass resolution, respectively.
$\sigma_{H1,2}$ was evaluated for each reconstructed Higgs boson
by using 31\%$/\sqrt{E_{\mathrm{jet}}}$, where $E_{\mathrm{jet}}$ is the jet energy.
Figure \ref{fig:nocut} shows the distribution of the sum of 
the two reconstructed Higgs boson masses 
for \nnhh~and background events.
With no selection cuts, 
the signal is swamped in huge number of background events.

To identify the signal events from the backgrounds, 
we applied the following selection cuts.
We required $\chi^{2} < 20$ and 95 GeV$< M_{H1,2} <$ 125 GeV
to select events, for which the Higgs bosons could be well reconstructed.
Then, since a $Z$ boson is missing in \nnhh~events,
we set the selection cut on the missing mass ($^{\mathrm{miss}}M$):
90 GeV $< ^{\mathrm{miss}}M < $ 170 GeV.

\begin{wrapfigure}{r}{0.55\columnwidth}
\centerline{\includegraphics[width=0.5\columnwidth]{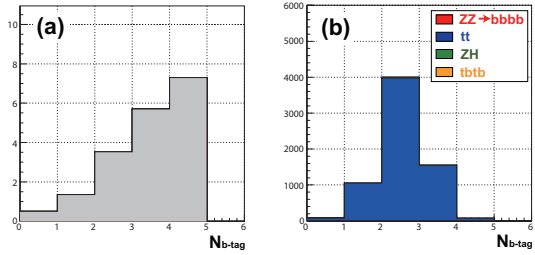}}
\caption{Distribution of the number of jets tagged as $b$-jets
after the selection cuts for \nnhh~(a) and backgrounds (b).}
\label{fig:nb}
\end{wrapfigure}

The angular distribution of the particles reconstructed as the Higgs bosons
has a peak at $\cos \theta = \pm 1$ for $ZZ$ events
whereas the distribution becomes more uniform in \nnhh~events. 
We applied the angular cut of $| \cos \theta_{\mathrm{H1,2}}| < 0.9$
to reject these $ZZ$ events.

The 4-jet events from $ZH$ events 
have small missing transverse momentum ($^{\mathrm{miss}} P_{\mathrm{T}}$),
which contaminate in the signal region.
For that reason, we required $^{\mathrm{miss}} P_{\mathrm{T}}$ above 50 GeV.

%\begin{wraptable}{l}{1.0\columnwidth}
\begin{table}[b]
\centerline{
\begin{tabular}{|l|rrrrr|} \hline
                & \nnhh       & $ZZ \to bbbb$ & $tt$      & $ZH$    & $tbtb$  \\ \hline
No cut          & 77.6        & 18,100        & 1,167,200 & 124,200 & 2,154 \\
$\chi^{2} < 20$ & 43.7        & 12,169        & 364,921   & 83,065  & 468 \\
95 GeV$< M_{H1,2} <$ 125 GeV & 29.5 & 387 & 70,557 & 8,570 & 82 \\ 
90 GeV$< ^{\mathrm{miss}}M <$ 170 GeV & 26.2 & 127 & 32,570 & 696  & 45 \\
$| \cos \theta_{H1,2}| < 0.9$ & 23.0 & 34.4 & 26,521 & 447 & 37 \\
$^{\mathrm{miss}} P_{\mathrm{T}} > 50$ GeV & 18.4 & 3.6 & 17,591 & 137 & 25 \\
$N_{\mathrm{lepton}} = 0$ & 17.8 & 3.6 & 6,708 & 37.3 & 9.7 \\
$N_{\mathrm{b-tag}} = 4$ & 7.3 & 1.8 & 65 & 0 & 2.4 \\   
\hline
\end{tabular}}
\caption{Cut statistics.}
\label{tb:cut_rate}
%\end{wraptable}
\end{table}

After the selection cuts so far, the dominant background was $tt$ events. 
The leptonic decay mode of $W$ from $t \to b W$ can be rejected 
by indentifying isolated charged leptons.
We define the energy deposit within 20 deg. around a track as $E_{20}$.
The isolated lepton track was defined to be a track with 
10 GeV $< E_{20} < \frac{2}{11} E_{\mathrm{trk}} - 1.8$ GeV,
where $E_{\mathrm{trk}}$ is energy of the lepton track.
We required the number of isolated lepton tracks ($N_{\mathrm{lepton}}$) to be zero.

Finally, the flavor tagging was applied.
We identified a jet as a $b$-jet, when it has 2 tracks 
with 3-sigma separation from the interaction point.
Figure \ref{fig:nb} shows the distribution of the number of jets tagged as $b$-jets
after the selection cuts ($N_{\mathrm{b-tag}}$).
Since the Higgs boson decays into $b\bar{b}$ with a 76\% branching ratio,
\nnhh~events have a peak at $N_{b\mathrm{-tag}} = 4$,
whereas $tt$ events have a peak at 2.
To reject the $tt$ events effectively, we selected events with $N_{b\mathrm{-tag}} = 4$.

\begin{wrapfigure}{r}{0.45\columnwidth}
\centerline{\includegraphics[width=0.4\columnwidth]{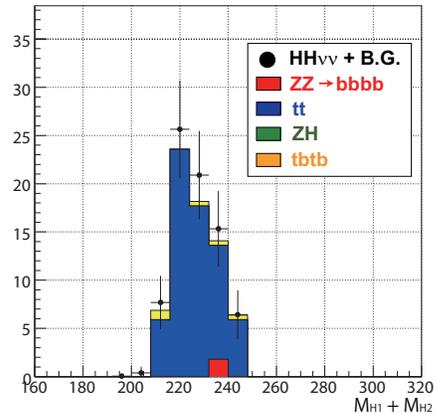}}
\caption{Distribution of the sum of the two reconstructed 
Higgs boson masses for \zhhnnhh~after all the selection cuts.}
\label{fig:bgsig}
\end{wrapfigure}

Figure \ref{fig:bgsig} shows the distribution of 
the sum of the two reconstructed Higgs masses for \zhhnnhh~after all the selection cuts.
We summarize the reduction rate by each selection cut in Table \ref{tb:cut_rate}.
Finally, we obtained 7.3 events for \nnhh~and 69.2 events for backgrounds.
This result corresponds to a signal significance of 0.9 ($= 7.3/\sqrt{69.2}$).
For observation of the $ZHH$ production, further background rejection, 
especially $tt$ events, is necessary.

%%%%%%%%%%%%%%%%%%%%%%%%%%%%%%%%%%
\section{Summary}
%%%%%%%%%%%%%%%%%%%%%%%%%%%%%%%%%%

The \zhhnnhh~process was studied to investigate 
the possibility of the trilinear Higgs self-coupling at the ILC.
In this study, we assumed the Higgs boson mass of 120 GeV,
$\sqrt{s} = 500$ GeV, and the integrated luminosity of 2 ab$^{-1}$.

After the selection cuts,
the signal significance of 0.9 was obtained.
The dominant background was $tt$ events.
To extract the information of \lhhh,
we must improve the background rejection.
In addition, 
$ZHH \to q \bar{q}HH$ will be studied to obtain certain signal significance.

%%%%%%%%%%%%%%%%%%%%%%%%%%%%%%%%%%
\section{Acknowledgments}
%%%%%%%%%%%%%%%%%%%%%%%%%%%%%%%%%%
The authors would like to thank all the members of the ILC physics subgroup
\cite{softg} for useful discussions. 
This study is supported in part by the Creative Scientific Research Grant
No. 18GS0202 of the Japan Society for Promotion of Science, and
Dean's Grant for Exploratory Research in Graduate School of Science of Tohoku University.

% ****************************************************************************
% BIBLIOGRAPHY AREA
% ****************************************************************************

\begin{footnotesize}
% IF YOU DO NOT USE BIBTEX, USE THE FOLLOWING SAMPLE SCHEME FOR THE REFERENCES
% ----------------------------------------------------------------------------

% ----------------------------------------------------------------------------

% IF YOU USE BIBTEX,
% - DELETE THE TEXT BETWEEN THE TWO ABOVE DASHED LINES
% - UNCOMMENT THE NEXT TWO LINES AND REPLACE 'Name_Of_Your_BibFile'

%\bibliographystyle{unsrt}
%\bibliography{Name_Of_Your_BibFile}
% example of Name_Of_Your_BibFile.bib
% @Article{Turcato:2006ch,
%      author    = "Turcato, M.",
%  collaboration = "ZEUS and H1",
%      title     = "Lepton flavour violation and charmonium physics at HERA",
%      journal   = "Nucl. Phys. Proc. Suppl.",
%      volume    = "162",
%      year      = "2006", 
%      pages     = "283-287",
%      SLACcitation  = "%%CITATION = NUPHZ,162,283;%%"
% }
% 
% @Unpublished{Gogitidze:2007du,
%      author    = "Gogitidze, N.",
%  collaboration = "H1", 
%      title     = "Prompt photons and particle momentum distributions at
%                   HERA", 
%      year      = "2007",
%      note    = "hep-ex/0701033",
%      SLACcitation  = "%%CITATION = HEP-EX 0701033;%%"
% }

\end{footnotesize}

% ****************************************************************************
% END OF BIBLIOGRAPHY AREA
% ****************************************************************************

\end{document}